\documentclass[preprint,12pt]{aastex}

\DeclareGraphicsExtensions{.eps}
\newcommand{\be}{\begin{equation}}
\newcommand{\ee}{\end{equation}}
\newcommand{\ba}{\begin{eqnarray}}
\newcommand{\ea}{\end{eqnarray}}
\newcommand{\bc}{\begin{center}}
\newcommand{\ec}{\end{center}}

\begin{document}

\title{
Luminous infrared galaxies  as plausible \\ $\gamma$-ray sources
for GLAST and IACTs}

\author{Diego F. Torres\altaffilmark{1}, Olaf
Reimer\altaffilmark{2}, Eva
Domingo-Santamar\'{\i}a\altaffilmark{3},
 \&
Seth W. Digel\altaffilmark{4}
 }

\altaffiltext{1}{Lawrence
Livermore National Laboratory, 7000 East Ave., L-413, Livermore,
CA 94550, E-mail: dtorres@igpp.ucllnl.org}
\altaffiltext{2}{Institut f\"ur Theoretische Physik,
Ruhr-Universit\"at Bochum, 44780, Germany. E-mail:
olr@tp4.ruhr-uni-bochum.de}
\altaffiltext{3}{Institut de F\'{\i}sica d'Altes
Energies (IFAE), Edifici C-n, Campus UAB, 08193 Bellaterra, Spain.
E-mail: domingo@ifae.es}
\altaffiltext{4}{W.W. Hansen Experimental Physics
Laboratory, Stanford University, Stanford,
                   CA 94305, E-mail: digel@stanford.edu}

\begin{abstract}
We argue that luminous infrared galaxies (LIGs) may constitute a
newly detectable population of $\gamma$-ray sources for the next
generation of ground and space-based high energy telescopes.
Additionally, we report for the first time upper limits on their
fluxes using data obtained with the EGRET telescope.
\end{abstract}

\section{Introduction}

Diffuse $\gamma$-ray emission from cosmic-ray (CR) interactions
with interstellar gas and photons is an important source of the
$\gamma$-ray luminosity of the universe, although the luminosities
of individual normal galaxies are relatively modest. Approximately
90\% of the high-energy $\gamma$-ray luminosity of the Milky Way
($\sim$1.3 $\times 10^6$ L$_\odot$, Strong, Moskalenko, \& Reimer
2000) is diffuse emission from CR interactions with interstellar
gas and photons (Hunter et al. 1997).  To date, the LMC is the
only external galaxy that has been detected in the light of its
diffuse $\gamma$-ray emission (Sreekumar et al. 1992).  Only upper
limits have been obtained for other local group galaxies (e.g.,
Pavlidou \& Fields 2001).  At 1 Mpc, for example, the flux of the
Milky Way would be approximately 2.5 $\times 10^{-8}$ photons
cm$^{-2}$ s$^{-1}$ ($>$100 MeV), well below the sensitivity of
past $\gamma$-ray missions. Although normal galaxies, or even
galaxies with fairly intense star formation such as the LMC, are
quite numerous, their distances make them very faint $\gamma$-ray
sources.

Luminous infrared galaxies (LIGs) are the dominant population of
extragalactic objects in the local universe ($z<0.3$) at
bolometric luminosities above $L > 10^{11}$ L$_\odot$. Some,
having $L_{\rm FIR} > 10^{12}$ L$_\odot$, are the most luminous
local objects (see Sanders \& Mirabel 1996 for a review). These
galaxies possess very large amounts of molecular gas (e.g.,
Sanders 1991; Downes et al. 1992; 1993; see below). Consequently,
they have large CO luminosities, but also a high value for the
ratio $L_{\rm FIR}/L_{\rm CO}$, both being about one order of
magnitude greater than for normal spirals. The latter implies,
based on star formation models, a greater star formation rate per
unit mass of gas. The molecular gas is often found concentrated in
the central regions of the galaxies, at densities orders of
magnitude larger than in Galactic giant molecular clouds. LIGs are
generally regarded as recent galaxy mergers in which much of the
gas of the colliding objects has fallen into a common center
(typically less than 1 kpc in extent), triggering a huge starburst
(e.g., Sanders et al. 1988, Melnick \& Mirabel 1990).\footnote{In
the case of Arp 302, however, the large CO luminosity comes from
an unusually large ($\sim 7 \times 10^{10}$ M$_\odot$) and
spatially extended amount of molecular gas, while it appears to
form stars at the Galactic rate (Lo et al. 1997). See below for
more discussion on this issue.}
Evidence indicates that the infrared luminosities of LIGs are
indeed due to starburst regions rather than enshrouded active
galactic nuclei (AGNs). In some cases, as in Arp 299, even when a
hidden AGN was observed, it cannot account for the whole FIR
luminosity (Della Cecca et al. 2002).

The large masses of dense interstellar gas and the enhanced
densities of supernova remnants and massive young stars suggest
that LIGs may have $\gamma$-ray luminosities orders of magnitude
greater than normal galaxies.  Using a simple criterion, we
explore here the prospects for detecting these objects with the
next generation of $\gamma$-ray telescopes.  A more detailed
theoretical model of $\gamma$-ray emission for an individual LIG
will be presented in a companion paper. (See Blom et al. 1997 and
Paglione et al. 1996 for examples of flux estimations from nearby
starburst galaxies.) Upper limits from existing EGRET data for the
fluxes of likely $\gamma$-ray--bright LIGs are also presented.

\section{Flux estimations and observability}

Neglecting possible CR density gradients within the interstellar
medium of a LIG, the hadronically-generated $\gamma$-ray number
luminosity (photons per unit of time) is (see, e.g., Torres et al.
2003c and references therein for details) $
I_{\gamma}(E_{\gamma})=\int n(r) q_{\gamma}(E_\gamma) dV =
({M}/{m_p}) {q_\gamma}, $ where $ r$ represents the position
within the interaction region $V$, $M$ is the mass of gas, $m_p$
is the proton mass, $n$ is the number density, and ${q_\gamma}$ is
the $\gamma$-ray emissivity (photons per unit of time per atom).
The $\gamma$-ray flux is then $F(>100\, {\rm MeV})=I_\gamma(>100
\, {\rm MeV})/4\pi {D_L}^2$, where $D_L$ is the luminosity
distance in a Friedman universe. In an appropriate scaling, $
F(>100\; {\rm MeV}) \sim 2.4 \times 10^{-9} ( {M}/{10^9 {\rm
M}_\odot}) ( {D_L}/{{\rm Mpc}} )^{-2} k \;{\rm photons\; cm^{-2}
s^{-1}}, $
 where $k$ is the
enhancement factor of CRs. The previous estimation assumes that $
k \sim {q_\gamma}/{q_{\gamma,\oplus}} \sim \epsilon /
\epsilon_\oplus$ with ${q_{\gamma,\oplus}} = 2.4 \times 10^{-25}$
photons s$^{-1}$ H-atom$^{-1}$ and $\epsilon_\oplus$ being the CR
density near Earth. The numerical factor already takes into
account the $\gamma$-ray emissivity from electron bremsstrahlung
(see, e.g., Pavlidou \& Fields 2003 and references
therein).\footnote{Note that $\gamma$-rays can also be produced by
inverse Compton interactions with the strong FIR field of the
galaxy. However, we shall disregard this contribution in favor of
the hadronic channel (between accelerated protons and diffuse
material of density $n$), which is a well justified approach above
100 MeV (see, e.g., Paglione et al. 1996 and references therein).
We also disregard additional hadronic production of high energy
$\gamma$-rays with matter in the winds of  stars (see, e.g.,
Romero \& Torres 2003; Torres et al. 2003). Both these effects
would improve the possibility for the galaxies to be detected. }
Note that $ F(>100\; {\rm MeV}) \sim 2.4 \times 10^{-9} \;{\rm
photons\; cm^{-2} s^{-1}} $ is approximately the GLAST satellite
sensitivity after 1 yr of all-sky survey.
%
%
A similar estimation can be made for the TeV flux expected from
these objects. V\"olk et al. (1996) found, $ F(>1 \; {\rm TeV})
\sim 1.7 \times 10^{-13} ( {E}/{{\rm TeV}})^{-1.1} ( {M}/{10^9
{\rm M}_\odot}) ( {D_L}/{{\rm Mpc}} )^{-2} k\; {\rm photons}\;
{\rm cm}^{-2} \;{\rm s}^{-1}$,  where a power law slope of 2.1 is
assumed for the CR spectrum. Here, $1\times 10^{-13}$ photons
cm$^{-2}$ s$^{-1}$ is the expected 5$\sigma$ flux sensitivity for
a 50 hr observation at small zenith angle at the new ground-based
imaging atmospheric \v{C}erenkov telescopes (IACTs). Then, those
galaxies that might appear in the new GeV catalogs might also
constitute new targets for the ground-based telescopes at higher
energies, provided their proton spectrum are sufficiently hard. In
addition, the signal-to-noise ratio in neutrino telescopes
(neutrinos will be unavoidably produced in hadronic interactions
leading to charged pions) can be approximately computed starting
from the $\gamma$-ray flux (see, e.g., Anchordoqui et al. 2003).
The result is that LIGs would be new candidate sources for ICECUBE
if they are detectable sources of TeV photons. Once produced,
photons (and of course also neutrinos) escape the FIR-dominant
field of the galaxy (which can be seen using the $\gamma\gamma$
and $\gamma p$ pair-production cross sections, Cox 1999, p.213ff,
and typical LIG parameters, e.g., the $< 1$ kpc-radius of the
central starburst where the $\gamma$-ray emission proceeds). In
addition, the small redshifts ($z < 0.05$) for the LIGs we are
considering make opacities due to processes with photons of the
CMB and IR-background negligible (see, e.g., Stecker 1971), 
so that once they escape the galaxy they may reach Earth
unscathed.

\section{A Plausible LIG--$\gamma$-ray connection}

LIGs not only  possess a large amount of molecular gas, but a
large fraction of it is at high density (e.g., Gao \& Solomon
2003a, b). This  makes them prone to star formation, and thus to
have significant CR enhancements. In nearby starburst galaxies,
like M82 and NGC253, the supernova rate is estimated to be at
least $\sim 0.1-0.3$ yr$^{-1}$ (Rieke et al. 1980), comparable to
the massive star formation rate (SFR), which is at least $\sim
0.1$ M$_\odot$ yr$^{-1}$ (Ulvestad et al. 1997, Forbes et al.
1993). The SFR in the LIGs  is 100--1000 times larger (e.g., Gao
\& Solomon 2003a, Fig. 6; Pasquali et al. 2003) and scales with
the amount of dense molecular gas (traced by the HCN line). V\"olk
et al. (1989) have calculated the CR energy density of M82 from
minimum energy considerations, assuming a proton/electron ratio of
100 as it is locally observed. They obtain a value $\epsilon_{\rm
M82}\sim 80$ eV cm$^{-3}$. Suchkov et al. (1993) estimate
$\epsilon_{\rm M82}\sim 170$ eV cm$^{-3}$ from the synchrotron
flux, and an upper limit of $\sim 10^4$ eV cm$^{-3}$ from the
observed density of molecular clouds. (The H$_{2}$ molecules are
destroyed by CRs and hence they dominate the mass of the clouds
only for a few hundred thousand years). It is natural to expect
that (the central regions of) LIGs have yet bigger enhancements,
comparable to the ratio between their SFRs and that of the Milky
Way. Interestingly, there is evidence for the existence of extreme
starbursts regions within LIGs  (see, e.g., Downes \& Solomon
1998). These, larger than giant molecular clouds but with
densities found only in small cloud cores, appear to be the most
outstanding star-forming regions in the local universe (each
representing about 1000 times as many OB stars as 30 Doradus).
The CR enhancement factor in these small but massive regions can
well exceed the average value for the galaxy. In Arp 220, for
instance, two such regions were discovered to contain about $ 2
\times 10^9$ M$_\odot$ (Downes \& Solomon 1998). If the local CR
enhancement there is significantly larger than the starburst
average, these extreme environments could be the main origin for
any $\gamma$-ray emission observed from this galaxy.

In order to establish the plausibility of the future detection of
LIGs in the $\gamma$-ray band, we consider the HCN survey just
presented by Gao \& Solomon (2003a,b). This survey is a systematic
observation (essentially, all galaxies with strong CO and IR
emission were chosen for HCN survey observations)  of 53 IR-bright
galaxies, including 20 LIGs with $L_{\rm FIR}>10^{11}$L$_\odot$, 7
with $L_{\rm FIR}>10^{12}$L$_\odot$, and more than a dozen of the
nearest normal spiral galaxies. It also includes a literature
compilation of data for another dozen IR-bright  objects. This is
the largest and most sensitive HCN survey (and thus of dense
interstellar mass) of galaxies to date. We have computed the
minimum average value of $k$ for which the $\gamma$-ray flux above
100 MeV is above 2.4 $\times 10^{-9}$ photons cm$^{-2}$ s$^{-1}$.
Luminosity distances used were those provided in the HCN survey,
assuming a Hubble parameter of $H_0$=75 km s$^{-1}$ Mpc$^{-1}$;
although, since redshifts are low enough, changes in the
cosmological model do not introduce significant changes in
distances.

Naively, the smaller the value of $\langle k\rangle$, the greater
the possibility for these Galaxies to appear as $\gamma$-ray
sources. All but three objects in the HCN survey require a value
$\langle k \rangle <10^3$ to be above the GLAST satellite
sensitivity, see Fig. 1 (left panel), where their luminosity
distance distribution is also shown in the inset. A typical case
for a plausible new GLAST source would be a LIG with $L_{\rm FIR}
\sim 10^{11}$L$_\odot$, $D_L \sim 10-100$ Mpc, $M($H$_2) \sim
10^{10}$M$_\odot$, and a CR enhancement of the order of 100. This
enhancement is an average value over the innermost central
starburst region, where most of the CO and HCN luminosity is
observed (see, e.g., Taniguchi \& Ohyama 1998, Gao \& Solomon
2003a,b), i.e., it considers that all molecular mass in that
region is illuminated by the enhanced CR spectrum. It is
reasonable to expect local variations from the average $k$,
particularly if extreme starbursts regions are separate from the
core, but yet it is a useful criterion for an observability study
(V\"olk et al. 1996).

The masses used to generate Fig. 1 assume the standard conversion
factors $X$ between CO or HCN luminosity and H$_2$ molecules (Gao
\& Solomon 2003a,b). We note, nevertheless, that several authors
presented the case for a reduction in these factors when applied
to powerful starburst regions, particularly with $L_{\rm FIR} >
10^{12}$L$_\odot$. Downes \& Solomon (1998) derived gas masses
from a model of radiative transfer, finding gas masses a factor of
$\sim 3-5$ lower than previous estimates. Solomon et al. (1997)
(see also Bryant \& Scoville 1999) concurred, but showed that even
after reducing $X$ by such amount, still the best estimations for
molecular masses in LIGs are $ (1.0\pm 0.3) \times 10^{10}$
M$_\odot$. Thus, we retain (as suggested in the HCN survey) all
estimations of mass using the standard conversion factors. (Note
that most LIGs in our sample have $L_{\rm FIR} < 10^{12}$L$_\odot$
anyway, and that an uncertainty of a factor of a few in $\langle k
\rangle$ would not affect the plausibility for detection in most
cases). Additionally, the HCN survey gives only H$_2$ molecular
mass; contributions to the full content of interstellar matter
other than H$_2$, e.g., by He, are not considered.


We have also considered the larger Pico dos Dias Survey (PDS,
Coziol et al. 1998), consisting of relatively nearby and luminous
galaxies selected in the FIR. PDS galaxies have a lower mean IR
luminosity $\log({\rm L}_{\rm IR}/{\rm L}_\odot) = 10.3 \pm 0.5$,
redshifts smaller than 0.1, and form a complete sample limited in
flux in the FIR at $2\times10^{-10}$ erg cm$^{-2}$ s$^{-1}$. CO or
other line measurements for most of the PDS galaxies are currently
unavailable. Thus, in order to get a first insight as to how many
PDS galaxies could appear as $\gamma$-ray sources for GLAST we
have applied the correlation between the $L_{\rm HCN}$ and $L_{\rm
IR}$, which is tight throughout three orders of magnitude (see
e.g., Solomon et al. 1992, Gao \& Solomon 2003a,b), to deduce
$L_{\rm HCN}$.\footnote{It is important to notice, while analyzing
the PDS sample, that the correlations derived in the HCN survey
entertain the IR luminosity, not the FIR luminosity. Then, one
needs to recompute --starting from the IRAS Catalog-- the IR
luminosities of all PDS starbursts when applying such
correlation.} We then use the correlation between $L_{\rm HCN}$
and $L_{\rm CO}$, which is tight for less luminous galaxies such
as those in the PDS (see Fig. 3 of Gao \& Solomon 2003a), to
estimate the total amount of molecular gas. 153 out of 203 PDS
galaxies (75\%) need an average enhancement $\langle  k \rangle
<500$ to appear as $\gamma$-ray sources in the next generation of
catalogs. In the middle panel of Fig. 1 we show the distribution
of CR enhancements required for GLAST detection as well as, in the
inset, the luminosity distances for all galaxies in the PDS. These
distributions appear different from those corresponding to HCN
galaxies. The reason is that the PDS galaxies are less IR-luminous
than the latter, contain less molecular mass, and thus require
larger enhancement factors (typically a factor of 3--5 larger than
a typical case in the HCN survey) to be detectable by GLAST.
Quantitatively, only 5\% of the PDS galaxies have star formation
rates larger than 100 M$_\odot$ yr$^{-1}$ (just 10 galaxies out of
203), the largest being 255 M$_\odot$ yr$^{-1}$. This has to be
compared with 20\% of the HCN galaxies having the same
characteristics, with the largest star formation rate being 660
M$_\odot$ yr$^{-1}$. Then, remarkably, even when PDS galaxies are
still IR-bright, we do not expect most of them to appear in the
GLAST catalog.

The reason why this is so is more clearly seen in the right panel
of Fig. 1.  There we show, for the HCN Galaxies, a plausible value
of the CR enhancement (obtained as the ratio between the SFR of
the galaxy and that of our Milky Way\footnote{The latter is also
obtained from the correlations given in the HCN survey, i.e., from
the Milky Way's HCN luminosity (e.g., Solomon et al. 1992, Wild \&
Eckart 2000) as SFR(MW)=$1.8M_{\rm dense}/(10^8$M$_\odot$)
M$_\odot$ yr$^{-1}$= $1.8 \times 10^{-7}(L_{HCN}/$ K km s$^{-1}$
pc$^{2}$) M$_\odot$ yr$^{-1}$. In any case, we note that the value
of $L_{HCN}$ for our Galaxy is uncertain and thus this panel
should be taken as indicative, not as a precise prediction of
$k$.}) versus the needed value of $k$ in order to make the galaxy
detectable by GLAST. Galaxies appearing above or around the line
of unit slope are prime candidates for detection. While a galaxy
with high $L_{HCN}/L_{CO}$ ratio (i.e., with a high mass fraction
of dense gas) will be a LIG (or a ULIG), the converse is not
always true (Gao \& Solomon 2003a). There are gas-rich galaxies
which are LIGs only because of the huge amount of molecular gas
they possess, not because they have most of it at high density
(and thus are undergoing a strong starburst phenomenon). In some
of these cases, while the value of enhancement needed for
detection might only be of  a few hundreds, the plausible value of
$k$ is much lower, since no strong star formation is
ongoing.\footnote{For example, in the HCN survey, there are a
group of 7 LIGs (out of 31) that are gas-rich (CO-luminous) but
have normal star formation efficiency $L_{IR}/L_{CO}$ (i.e.,
$L_{HCN}/L_{CO} < 0.06$). Some examples are NGC 1144, Mrk 1027,
NGC 6701, and Arp 55. They are using the huge molecular mass they
have in creating stars at a normal SFR. }In the context of
$\gamma$-ray observability, GLAST will detect those galaxies that,
being close enough, not only shine in the FIR but that do so {\it
because} of their active strong star formation processes.

In any case, since there are good candidates for detection in both
surveys we suggest that both the HCN and PDS samples be taken into
account when planning population analyses with the next generation
of catalogs of point-like $\gamma$-ray sources.


\begin{figure*}[t]
\centering
\includegraphics[width=0.92\textwidth,height=6.5cm]{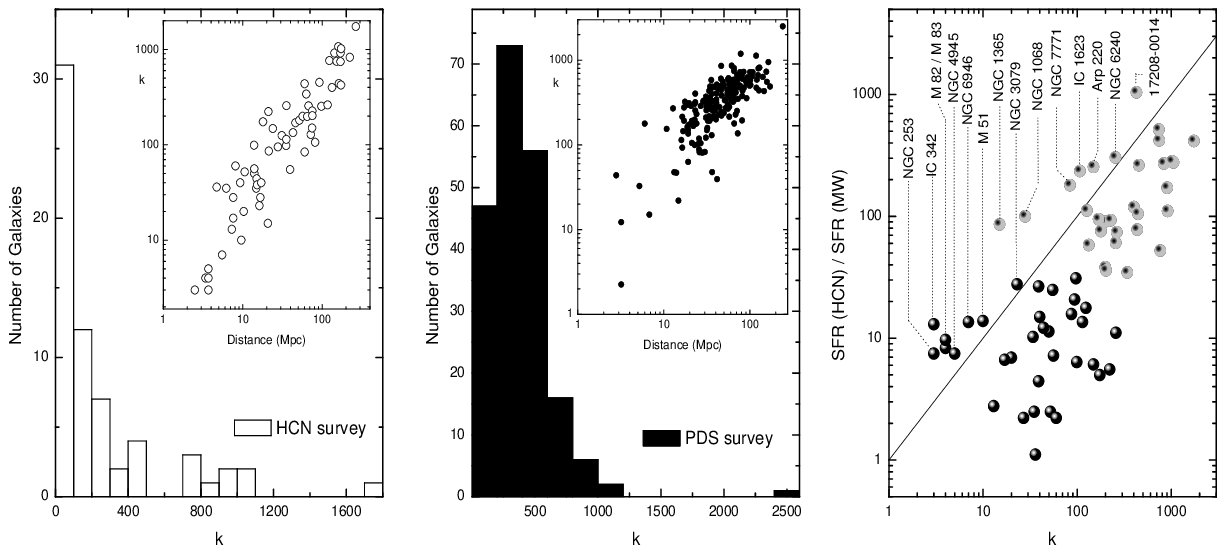}
\caption{Distribution of luminosity distances and minimum average
CR enhancements needed for galaxies in the HCN (left) and Pico Dos
Dias (middle) surveys to appear as $\gamma$-ray sources for GLAST
and the new ground based \v{C}erenkov telescopes. Right: Plausible
values of enhancements of the HCN galaxies obtained as the ratio
between the SFR of each galaxy and that of the Milky Way versus
the needed one for them to be detectable by GLAST. LIGs (less
luminous galaxies) are shown as white points (black) points. See
text for discussion.}
\end{figure*}



\section{The EGRET data}

We consider a sample of galaxies, a sub-sample of the HCN survey
also observed with the Owens Valley Radio Observatory (Scoville et
al. 1991) and the IRAM interferometer (Downes \& Solomon 1998, see
also Taniguchi and Ohyama 1998), for an investigation regarding
the presently available $\gamma$-ray data. These galaxies are
shown in Table 1. There, distances are obtained from their NED
redshifts under the assumption that $H_0=75$ km s$^{-1}$
Mpc$^{-1}$ and $q_0=0.5$. For their molecular mass content, IR,
and CO luminosities see the HCN survey, {\it op. cit.}

Data from the Energetic Gamma Ray Experiment Telescope (EGRET)
aboard the Compton Gamma Ray Observatory (CGRO) at energies above
100 MeV were co-added and analyzed at the respective locations of
the candidate objects. We present 95\% confidence upper limits for
the energy selection E $> 100$ MeV. They were obtained by a
likelihood analysis (Mattox et al. 1996) from data throughout the
CGRO observation cycles 1 to 9. The established EGRET sources from
the 3EG catalog (Hartman et al. 1999) were included in the
background model at the fluxes estimated by the likelihood
algorithm. None of the seventeen LIGs we investigated have been
detected, which is, in fact, consistent with the level of flux
expected from LIGs (i.e., fluxes above GLAST sensitivity but below
the EGRET one). The likelihood test statistic, TS, was maximal for
NGC2146, corresponding to a statistical significance of
1.9$\sigma$.
Upper limits on the $\gamma$-ray fluxes were obtained from the
likelihood analysis. Table 1 lists these upper limits in units of
$10^{-8}$ cm$^{-2}$ s$^{-1}$.
Two of the candidate sources have significant
PSF-overlap with known EGRET sources: NGC3079 is $\sim$1.6 degrees
away from 3EG~J0952+5501, and NGC520 is $\sim$1.8 degrees away
from 3EG~J0118+0248. The former is a high-probability blazar-class
AGN identification (0954+556); the latter, a tentative AGN
identification (0119+041, a FSRQ, as suggested in Hartman et al.
1999, or 0113+0222, a narrow-line radio galaxy at z=0.05 as
proposed in Sowards-Emmerd et al. 2003). In both cases, the flux
upper limit has been determined under full modeling and
subtraction of the respective EGRET source. Except in cases of a
source location near the galactic plane (NGC2146), the upper
limits determined for the fluxes are consistently on the order of
a few $10^{-8}$ cm$^{-2}$ s$^{-1}$.

\section{Concluding remarks}

There are several LIGs (among them the paradigmatic Arp 220, which
we shall analyze in detail elsewhere) for which reasonable values
of CR enhancements, comparable to, or lower than, the ratio
between their SFR and the Milky Way's, can provide a $\gamma$-ray
flux above GLAST sensitivity, and if the CR spectrum is
sufficiently hard, also above the sensitivities of the new
\v{C}erenkov telescopes. Some LIGs are then most likely to appear
as new point-like $\gamma$-ray sources. Even when it is natural to
expect that a LIG will emit copious $\gamma$-rays, only the more
gaseous, nearby, and CR-enhanced galaxies are the ones which could
be detected as point sources.
Note that only non-variable $\gamma$-ray sources can be ascribed
to LIGs. Variability indices (Torres et al. 2001, Nolan et al.
2003) can then play a role in the acceptance or rejection of
possible counterparts of  LIGs. The EGRET instrument was unable to
detect LIGs, at least the small sample explored in this Letter.
Flux upper limits at the source locations were imposed from EGRET
data which are consistent with the expected level of $\gamma$-ray
emission from LIGs.

The  work of DFT was performed under the auspices of the U.S.
Department of Energy (NNSA) by UC's  LLNL under contract No.
W-7405-Eng-48. Support for OR by the Bundesministerium f\"ur
Bildung und Forschung through DLR, grant 50 QV 0002 is gratefully
acknowledged. ED-S acknowledges the Ministry of Science and
Technology of Spain for financial support and the IGPP/LLNL for
hospitality. We acknowledge Y. Gao, G. Romero, and the Referee, V.
Pavlidou, for their insightful remarks.


\begin{table*}[t]
\caption{Upper limits (2$\sigma$) on the $\gamma$-ray flux and number luminosity from nearby LIGs. }
\begin{center}
\begin{tabular}{lccc} \hline
Name                 & $D_L$      & $F_{>100\, {\rm MeV} }^{\rm EGRET}$   & $I_{>100\, {\rm MeV} }^{\rm EGRET}$\\
                     & [Mpc]      & [photons cm$^{-2}$ s$^{-1}$] & [photons s$^{-1}$]\\
\hline
NGC3079              & 15        &  $<$4.4 $\times 10^{-8}$ & $<$1.2 $\times 10^{45}$  \\ 
NGC1068              & 15        &  $<$3.6 $\times 10^{-8}$ & $<$9.6 $\times 10^{44}$  \\
NGC2146              & 20        &  $<$9.7 $\times 10^{-8}$ & $<$4.6 $\times 10^{45}$  \\ 
NGC4038/9            & 22        &  $<$3.7 $\times 10^{-8}$ & $<$2.1 $\times 10^{45}$  \\
NGC520               & 29        &  $<$4.6 $\times 10^{-8}$ & $<$4.6 $\times 10^{45}$  \\
IC694                & 41        &  $<$2.2 $\times 10^{-8}$ & $<$4.4 $\times 10^{45}$  \\
Zw049.057            & 52        &  $<$6.9 $\times 10^{-8}$ & $<$2.2 $\times 10^{46}$  \\
NGC1614              & 64        &  $<$5.0 $\times 10^{-8}$ & $<$2.4 $\times 10^{46}$  \\
NGC7469              & 65        &  $<$3.2 $\times 10^{-8}$ & $<$1.6 $\times 10^{46}$  \\
NGC828               & 72        &  $<$6.1 $\times 10^{-8}$ & $<$3.7 $\times 10^{46}$  \\
Arp220               & 72        &  $<$6.1 $\times 10^{-8}$ & $<$3.7 $\times 10^{46}$  \\
VV114                & 80        &  $<$3.9 $\times 10^{-8}$ & $<$2.9 $\times 10^{46}$  \\
Arp193               & 94        &  $<$5.2 $\times 10^{-8}$ & $<$5.4 $\times 10^{46}$  \\
NGC6240              & 98        &  $<$6.4 $\times 10^{-8}$ & $<$7.3 $\times 10^{46}$  \\
Mrk273               & 152       &  $<$2.3 $\times 10^{-8}$ & $<$6.3 $\times 10^{46}$  \\
IRAS 17208$-$0014    & 173       &  $<$7.5 $\times 10^{-8}$ & $<$2.6 $\times 10^{47}$  \\
VIIZw31              & 217       &  $<$3.2 $\times 10^{-8}$ & $<$1.7 $\times 10^{47}$  \\
\hline \hline
\end{tabular}
\end{center}
\end{table*}



\end{document}